\documentclass[conference]{IEEEtran}
\IEEEoverridecommandlockouts

\usepackage{amsmath,amsfonts}
\usepackage{array}
\usepackage[caption=false,font=normalsize,labelfont=sf,textfont=sf]{subfig}
\usepackage{textcomp}
\usepackage{stfloats}
\usepackage{url}
\usepackage{verbatim}
\usepackage{graphicx}
\usepackage{cite}
\usepackage[table,xcdraw]{xcolor}
\usepackage{pifont}
\usepackage{algorithm}  
\usepackage{algpseudocode}  
\usepackage{multirow}
\usepackage{booktabs}
\usepackage{caption}
\usepackage{flushend}
\usepackage{makecell}

\usepackage{titlesec}
\titlespacing{\section}{3pt}{3.0pt plus 0.0pt minus 1.0pt}{3pt plus 1.0pt minus 1.0pt}
\titlespacing{\subsection}{3pt}{3pt plus 0.0pt minus 1.0pt}{3pt plus 1.0pt minus 1.0pt}

\begin{document}
\bstctlcite{IEEEexample:BSTcontrol}

\title{Michscan: Black-Box Neural Network Integrity Checking at Runtime Through Power Analysis
}

\author{%
Robi Paul and Michael Zuzak\\
\IEEEauthorblockA{Rochester Institute of Technology, Rochester, NY USA\\
\{rp7248, mjzeec\}@rit.edu}
}


\maketitle

\begin{abstract}
As neural networks are increasingly used for critical decision-making tasks, the threat of integrity attacks, where an adversary maliciously alters a model, has become a significant security and safety concern. These concerns are compounded by the use of licensed models, where end-users purchase third-party models with only \textit{black-box} access to protect model intellectual property (IP). In such scenarios, conventional approaches to verify model integrity require knowledge of model parameters or cooperative model owners. To address this challenge, we propose Michscan, a methodology leveraging power analysis to verify the integrity of black-box TinyML neural networks designed for resource-constrained devices. Michscan is based on the observation that modifications to model parameters impact the instantaneous power consumption of the device. We leverage this observation to develop a runtime model integrity-checking methodology that employs correlational power analysis using a golden template or signature to mathematically quantify the likelihood of model integrity violations at runtime through the Mann-Whitney U-Test. Michscan operates in a black-box environment and does not require a cooperative or trustworthy model owner. We evaluated Michscan using an STM32F303RC microcontroller with an ARM Cortex-M4 running four TinyML models in the presence of three model integrity violations. Michscan successfully detected all integrity violations at runtime using power data from five inferences. All detected violations had a negligible probability ($P < 10^{-5}\%$) of being produced from an unmodified model (i.e., false positive).   
\end{abstract}


\section{Introduction}
Due to increasing demands for efficiency, autonomy, and privacy, neural networks are frequently deployed on edge devices \cite{abadade2023comprehensive}. This trend has driven the development of \textit{TinyML} models, designed to operate in ultra-low-power and resource-constrained environments \cite{banbury2020benchmarking}. These models are often trained on large, proprietary datasets to achieve robust and accurate performance before being compressed for deployment on limited-resource devices \cite{zaidi2022unlocking}. This has led to research exploring the commercialization of models, where end-users purchase or license pre-trained models \cite{leroux2022tinymlops}. To protect intellectual property, model owners typically provide these models in a \textit{black-box} format, hiding internal parameters from the end-user.

The strong performance and robustness of neural networks have driven their broad adoption in critical decision-making tasks, such as autonomous driving, anomaly detection, and predictive maintenance \cite{kallimani2023tinyml}. This adoption spans from cloud-based machine learning as a service (MLaaS) to highly resource-constrained edge devices like remote sensors, wearables, and medical devices. Particularly in edge scenarios, where devices operate in potentially untrusted environments, neural networks become targets for integrity attacks, where the model is altered for adversarial goals \cite{gu2019badnets, guo2020fine}. We can consider these attacks in three broad categories: 1) Trojan attacks \cite{gu2019badnets}, 2) data poisoning-based attacks \cite{kravchik2021poisoning}, and 3) fault injection attacks \cite{liu2017fault, farmani2021rhat}. Trojan attacks, such as BadNets \cite{gu2019badnets}, are designed to cause incorrect classifications for specific stealthy inputs. Poisoning attacks involve fine-tuning the original model with malicious data to degrade performance or cause misclassification for specific inputs \cite{kravchik2021poisoning}. Fault injection attacks, such as the Bit-Flip Attack \cite{rakin2019bit}, aim to induce errors in the model to degrade or modify model behavior. Given the safety and security-critical applications where TinyML models are deployed, it is important that the integrity violations caused by such attacks are detected at runtime.  

A sizable body of prior research has already explored integrity checking mechanisms for neural networks \cite{kallimani2023tinyml, gu2019badnets, guo2020fine, gu2019badnets, kravchik2021poisoning, liu2017fault, farmani2021rhat, kravchik2021poisoning, rakin2019bit, luo2019functional, botta2021neunac, aramoon2021aid, he2019sensitive}. However, these works generally require access to a model and its internal parameters \cite{luo2019functional, shehab2018secure, botta2021neunac, anandakumar2022rethinking, zhang2012identification, chen2019deepattest, stern2018emforced}. In some cases, to facilitate a checking mechanism, the model itself must even be modified \cite{botta2021neunac, chen2019deepattest, stern2018emforced}. In a licensing scenario, the model owner may provide only black-box access to the end-user to protect critical model IP \cite{leroux2022tinymlops}. In such cases, the end-user has to trust the model owner, and the model owner must be willing to perform model modifications \cite{botta2021neunac} or generate test cases that may leak key model details \cite{aramoon2021aid, he2019sensitive}. Hence, while prior research has developed effective integrity-checking mechanisms, they inherently require either access to model parameters or a cooperative and trustworthy model owner. In a licensing scenario, this may not be possible, opening a model up to integrity attacks. This leads to our primary goal: \textit{\textbf{to develop a mathematically robust runtime integrity-checking mechanism for TinyML models without requiring trust or cooperation from the model owner.}}

\subsection{Contributions}

In this work, we propose Michscan, a runtime model integrity-checking mechanism for TinyML models deployed in resource-constrained edge devices. Michscan, which stands for ``Model Integrity CHecking through power Side Channel ANalysis", is designed to operate in a black-box setting without requiring trust or cooperation by the model owner. To perform runtime integrity checking, Michscan employs a Mann-Whitney U-Test to quantify the probability that an integrity violation has occurred based on the the instantaneous power consumption of a device during model inference.
The contributions of the work are summarized below:


\begin{enumerate}
\item We demonstrate that TinyML model integrity violations can be reliably detected through power analysis.
\item We develop Michscan, a runtime model integrity-checking mechanism that uses correlational power analysis to detect integrity violations in black-box TinyML models deployed in resource-constrained devices. Unlike prior art, Michscan does not require trust in or participation by the model owner to detect violations.
\item We develop a mathematically robust detection mechanism for Michscan based on Hypothesis Testing (Mann-Whitney U-Test). This mechanism provides quantifiable detection confidence for integrity violations and can be tuned by the end user to meet preferred detection goals.
\item We evaluate the performance of Michscan in an STM32F303RC MCU (ARM Cortex-M4) running four TinyML models in the presence of three integrity attacks: Trojan, poisoning, and fault injection. Across 1600 evaluated test cases, Michscan successfully detected all integrity violations $(P_{th} < 10^{-5})$ with no observed false positives at runtime using power data from only 5 model inferences.
\end{enumerate}

\section{Preliminaries}
\subsection{Integrity Attacks on Neural Networks}
\label{sec:attack}

Integrity attacks on neural networks aim to modify model parameters for adversarial goals, such as performance degradation or targeted misclassification. A large body of prior work has explored such attacks \cite{gu2017badnets, liu2018trojaning, kravchik2021poisoning, kurita2020weight, rakin2019bit, farmani2021rhat}. For this study, we broadly classify integrity attacks into 3 categories.

\textbf{Trojan Attacks}: These attacks cause incorrect classifications for a model when a specific \textit{trigger} input is received while allowing the model to function normally for other tasks. Trigger inputs often aim for stealth, relying on either uncommon or imperceptible changes to avoid detection. Examples of such attacks include BadNets \cite{gu2017badnets} and Neural Trojans \cite{liu2018trojaning}. The attacker typically retrains the model with a malicious dataset for these attacks. This process produces a new model with different parameters from the original, unattacked model. 

\textbf{Poisoning}: These attacks generally retrain or fine-tune a model using a malicious dataset to degrade performance for either a specific class or all classes \cite{kravchik2021poisoning, kurita2020weight}. To maintain stealth, attackers often fine-tune only specific layers when executing these attacks. Depending on the position of the targeted layer, the subsequent layers will diverge from those of the unattacked model.

\textbf{Fault Injection}: These attacks exploit physical aspects of hardware to either inject errors or influence model behavior. Examples of such attacks include modifying system temperature or voltage \cite{qui2020voltjockey} or creating electromagnetic disturbances to inject errors into memory \cite{sun2023lightning}. Alternative methods of fault injection, such as the Bit-Flip Attack \cite{rakin2019bit}, or RHAT \cite{farmani2021rhat}, have also been explored to induce model integrity violations. 

\subsection{Integrity Checking in Neural Networks}

A large body of work has explored the detection of integrity violations in neural networks. We broadly classify this work by degree of access, namely 1) white-box \cite{shehab2018secure, botta2021neunac, anandakumar2022rethinking, zhang2012identification}, 2) grey-box \cite{he2019sensitive}, and 3) limited black-box \cite{aramoon2021aid}. 

\textbf{White-Box:} These techniques require full access to model parameters and architecture to detect integrity violations. Prominent approaches in this category include watermarking \cite{chen2019deepmarks, shehab2018secure, botta2021neunac} and fingerprinting \cite{chen2019deepattest, stern2018emforced}. In such methods, the model owner is typically responsible for modifying the model with hard-to-replicate or hard-to-detect metadata or signatures (i.e., watermarks or fingerprints). Initially, such techniques were used for rights management \cite{anandakumar2022rethinking, zhang2012identification}. However, in cases of fragile watermarks \cite{shehab2018secure, botta2021neunac}, model re-training corrupts embedded signatures, enabling integrity checking. 


\textbf{Grey-Box:} These techniques assume users have limited access to a model's internal values, namely the ability to observe inputs to the final fully connected layer \cite{jo2024exploring, he2019sensitive}. For instance, prior work proposes \textit{sensitive samples} \cite{he2019sensitive}, which are owner-generated inputs outside of a model's training data. To detect model manipulations, a user can apply sensitive samples as inputs to the model and observe the response of the final fully connected layer. Any changes in the inputs to the final fully connected layer indicate an integrity violation within the model. This is because any alteration in model parameters in a prior layer can be observed at the input to the final layer. 



\textbf{Limited Black-Box:} These techniques can be used in a scenario where the end-user has no access to model parameters \cite{chen2019deepinspect, aramoon2021aid}. However, they require a cooperative model owner to generate the integrity-checking mechanism. For example, AID \cite{aramoon2021aid} performs model integrity checking using generated test inputs that reside near the decision boundary. While these test inputs can be applied by an end-user to perform integrity checking, they require white-box access to a model to generate them. Additionally, because these test inputs reside near the decision boundary, they may leak key details of the model. Hence, while AID can be deployed in a black-box scenario, it requires a white-box model owner for support. 

While these techniques all adopt different strategies to detect integrity violations, \textbf{they all assume that there is trust and cooperation between the owner and the user of the model}. However, in a licensing scenario, the model owner may provide only black-box access to the end-user to protect critical model IP \cite{leroux2022tinymlops}. In this case, the end-user has to trust the model owner, and the model owner must be willing to perform model modifications \cite{zhao2021veriml, chen2019deepattest, darvish2019deepsigns} or generate test cases that may leak key model details \cite{aramoon2021aid, he2019sensitive} to the end user to facilitate integrity checking. Hence, while prior research has developed several effective integrity-checking mechanisms, they inherently require a cooperative and trustworthy model owner. In a licensing scenario, this may not be possible. Therefore, prior work leaves the user vulnerable to an uncooperative or untrustworthy model owner who could either 1) refuse to generate the checking mechanism out of fear of leaking model IP or 2) manipulate the integrity checking mechanism to conceal their own manipulation. This significant gap in ensuring model integrity in licensing scenarios motivates our research. 

\subsection{Mann-Whitney U-Test}
\label{sec:mann}
The Mann-Whitney U-Test is a non-parametric hypothesis test that determines whether there is a statistically significant difference between the distributions of two groups \cite{nachar2008mann}. Because only the ranks of points in each group are compared, no assumption is made regarding the shape of the underlying distribution that produced each group (e.g., normality). This makes the Mann-Whitney U-Test an alternative to Z-test or T-test, which require data to be sampled from Gaussian distributions. In this work, we have adopted the Mann-Whitney U-Test for analysis because the characteristics of the underlying distribution being sampled are not fully understood. The test has three necessary conditions: 1) independence of observations, 2) continuous data, and 3) equal shape/spread. Additionally, for statistical validity, the test requires a minimum sample size of $n_{RA} \geq 5$ \cite{nachar2008mann}. We discuss each of these assumptions in greater detail with respect to the problem addressed in this work in Sec. \ref{sec: Runtime_need}. The equation for a Mann-Whitney U-Test is:
\begin{equation}
U_1 = n_1 n_2 + \dfrac{n_1(n_1 + 1)}{2} - R_1,
\end{equation}
\begin{equation}
U_2 = n_1 n_2 + \dfrac{n_2(n_2 + 1)}{2} - R_2, 
\end{equation}
\begin{equation}
n_1,n_2 \geq 5
\end{equation}
where $U_{x}$ is the U statistic used to assess the null hypothesis for group $x=\{1,2\}$, $n_x$ is the sample size of group $x=\{1,2\}$, and $R_x$ is the sum of the ranks assigned to the observations in group $x=\{1,2\}$.

The U statistics generated by the test correspond to a probability value, which can be looked up in an Mann-Whitney Table \cite{nachar2008mann}. If the two tested groups are sampled from distributions with equal shape/spread, as is done by this work, this probability value corresponds to whether the two groups are sampled from distributions with the same median. This serves as the null hypothesis. If the probability value is below a user-defined threshold, the test rejects the null hypothesis in favor of the alternative hypothesis, indicating a statistically significant difference between tested groups.


\subsection{Power Analysis for Anomaly Detection}

The instantaneous power consumption of a circuit, measured through supply voltage drop, varies based on both data and control flow \cite{brier2004correlation, kocher1999differential}. This is shown by a large body of work that uses power monitoring to reveal detailed data and control flow information, such as cryptographic keys \cite{kocher1999differential, brier2004correlation}. This is possible because the shared power distribution network (PDN) supplies power to all system components. The PDN aims to provide constant voltage despite varying current demands. However, maintaining a truly constant voltage is infeasible. Switching activities cause transient voltage drops in the PDN \cite{pant2008design}. These voltage drops can be measured, modeled, and used to infer information about the circuit's data and control flow \cite{brier2004correlation, kocher1999differential}. Prior work shows the feasibility of using power signature analysis \cite{laughman2003power} and other power analysis for anomaly detection, including program identification \cite{delimitrou2017bolt, greveler2012multimedia}, malware detection \cite{kim2008detecting, ding2020deeppower, zhang2023trustguard}, and Trojan detection \cite{wang2011power, shende2016side}.

\subsection{Threat Model}
\label{sec:4_point}

We consider a scenario where a TinyML model operates on a resource-constrained microcontroller unit (MCU). This is representative of an edge-deployed device, such as a remote sensor, wearable, or medical device, where an adversary has physical and/or remote access and can violate the integrity of the device post-deployment. Because these devices operate in an untrusted environment, they are particularly vulnerable to integrity violations. Additionally, due to their resource-constrained nature, we assume these devices cannot co-locate multiple applications simultaneously. The considered attacker can only gain access of the device running the model after it has been deployed to an untrusted environment and aims to compromise the model's integrity by modifying its parameters or architecture. Attacks such as Trojans \cite{gu2019badnets}, poisoning \cite{kravchik2021poisoning}, and fault injection \cite{liu2017fault} are considered in-scope \cite{gu2017badnets, liu2018trojaning, kravchik2021poisoning, kurita2020weight, rakin2019bit, farmani2021rhat}. The goal of such attacks can range from targeted misclassification (e.g., to bypass access controls) or performance degradation (e.g., to prevent detection of malicious behavior).

The end-user's objective is to maintain the model's integrity by identifying any changes at runtime. Unlike previous approaches, we assume that the defender has neither access to nor control over the model's parameters, and the model owner is unwilling to modify the model. This situation reflects a licensing scenario \cite{leroux2022tinymlops} where the model owner is reluctant to produce sensitive samples, fearing the leakage of decision boundary information, or to modify the model to include a fingerprint for integrity verification. Consequently, prior work that relies on accessing the model parameters \cite{chen2019deepattest, he2019sensitive} or a cooperative owner \cite{zhao2021veriml, chen2019deepattest, darvish2019deepsigns, aramoon2021aid, he2019sensitive} are not feasible. 

We define a successful defense strategy as one that can detect model integrity violations at runtime without requiring alterations to the model itself, cooperation from the model owner, or disclosure of the model's internal details to the end user. Our approach relies on establishing a golden template of the model's power consumption profile, which must be created in a trusted environment pre-deployment of the model free of integrity violations. To assess the efficacy of integrity checking, we adopt four success metrics established in prior work for evaluating integrity-checking solutions \cite{aramoon2021aid}, which are defined below.


\begin{enumerate}
\item \textbf{Effectiveness:} The probability of detecting changes to model parameters.
\item \textbf{Efficiency:} The performance overhead associated with the integrity checking approach.
\item \textbf{Reliability:} The probability of improper classification of integrity violations (i.e., false positive/negative).
\item \textbf{Generalizability:} The ability to generalize between model architecture, dataset, and parameters.
\end{enumerate}

\begin{table}[!b]
    \centering
    \caption{Access levels and protection features from commercial MLaaS providers.}
    \begin{tabular}{l|c c}
    \toprule
    \textbf{Service} & \textbf{Access Level} &\textbf{Current Protection}  \\
    \midrule
       AWS SageMaker \cite{aws-sagemaker}       & Black-box   & \makecell{Model registry,\\IAM roles,\\Container verification} \\ \addlinespace
       Azure ML Studio \cite{azure-ml}      & Black-box   & \makecell{Model signing,\\Role-based access,\\Audit trails} \\ \addlinespace
       Google Cloud AutoML \cite{google-automl} & Black-box   & \makecell{Binary authorization,\\Container security} \\ \addlinespace
       Amazon Rekognition \cite{amazon-rekognition}  & Black-box   & \makecell{API security,\\Version control}\\ \addlinespace
       Microsoft Cognitive \cite{microsoft-cognitive} & Black-box   & \makecell{Container validation,\\Access control}\\
    \bottomrule
    \end{tabular}
    \label{tab:mlaas}
\end{table}

\subsection{Commercial MLaaS: Runtime Integrity Gap}

Major cloud providers, including AWS SageMaker \cite{aws-sagemaker}, Azure ML Studio \cite{azure-ml}, and Google Cloud AutoML \cite{google-automl}, offer machine learning as a service (MLaaS) platforms with robust security mechanisms. A list of the security protections offered by each provider is aggregated in Tab. \ref{tab:mlaas}. These protections primarily consider pre-deployment and initial model verification, creating a critical gap in runtime model integrity checking. While Michscan cannot be implemented in its current form on these platforms due to the inability to access power traces from remote devices, if voltage/power sensors with sufficient granularity were made available through the provider, users could implement Michscan for runtime integrity checking.

\section{Motivation \& Problem Formulation}
\label{sec:motivation}


The increasing use of TinyML for critical decision-making tasks necessitates the development of a robust model integrity-checking mechanism. With the ongoing commercialization of machine learning, end users may only receive black-box copies of licensed TinyML models \cite{leroux2022tinymlops}, and the model owners may be reluctant to modify a model \cite{das2023psc} or provide sensitive samples \cite{he2019sensitive} due to concerns about information leakage. Therefore, there is a strong need to develop a model integrity-checking mechanism that functions in a \textit{black-box} environment with no cooperation required by the model owner. In this work, we explore the use of power analysis to assess the integrity of TinyML models in a black-box setting.

Power analysis offers insight into a processor's internal operations during model inference due to the relation between the processor's power consumption and its data/control flow \cite{brier2004correlation}. Any model modification (i.e., integrity violation) will alter either the data or control flow of the processor, impacting power consumption. Thus, power analysis can detect integrity violations. Such an approach functions in a black-box setting and is agnostic to the processor hardware, model, attack type, and data modality.

At the core of this approach is the assumption that two TinyML models running on an MCU can be reliably differentiated through power analysis. To demonstrate that this is indeed the case, let us consider the motivational scenario where an STM32F303RC (ARM Cortex-M4) is running a ResNet20 model trained on the CIFAR-10 dataset from the TinyML Perf benchmark \cite{banbury2021mlperf}. To simulate an integrity violation, we have also generated a poisoned version of this model by fine-tuning the model with incorrect training data. To monitor the power consumption of the MCU, the voltage drop across a $10\Omega$ resistive shunt is measured. To evaluate the feasibility of integrity checking, we collected power measurements of the MCU performing inference for a specific, randomly selected input using both the baseline and poisoned ResNet20 model.

\begin{figure}[t]
  \centering
  \includegraphics[width=\linewidth]{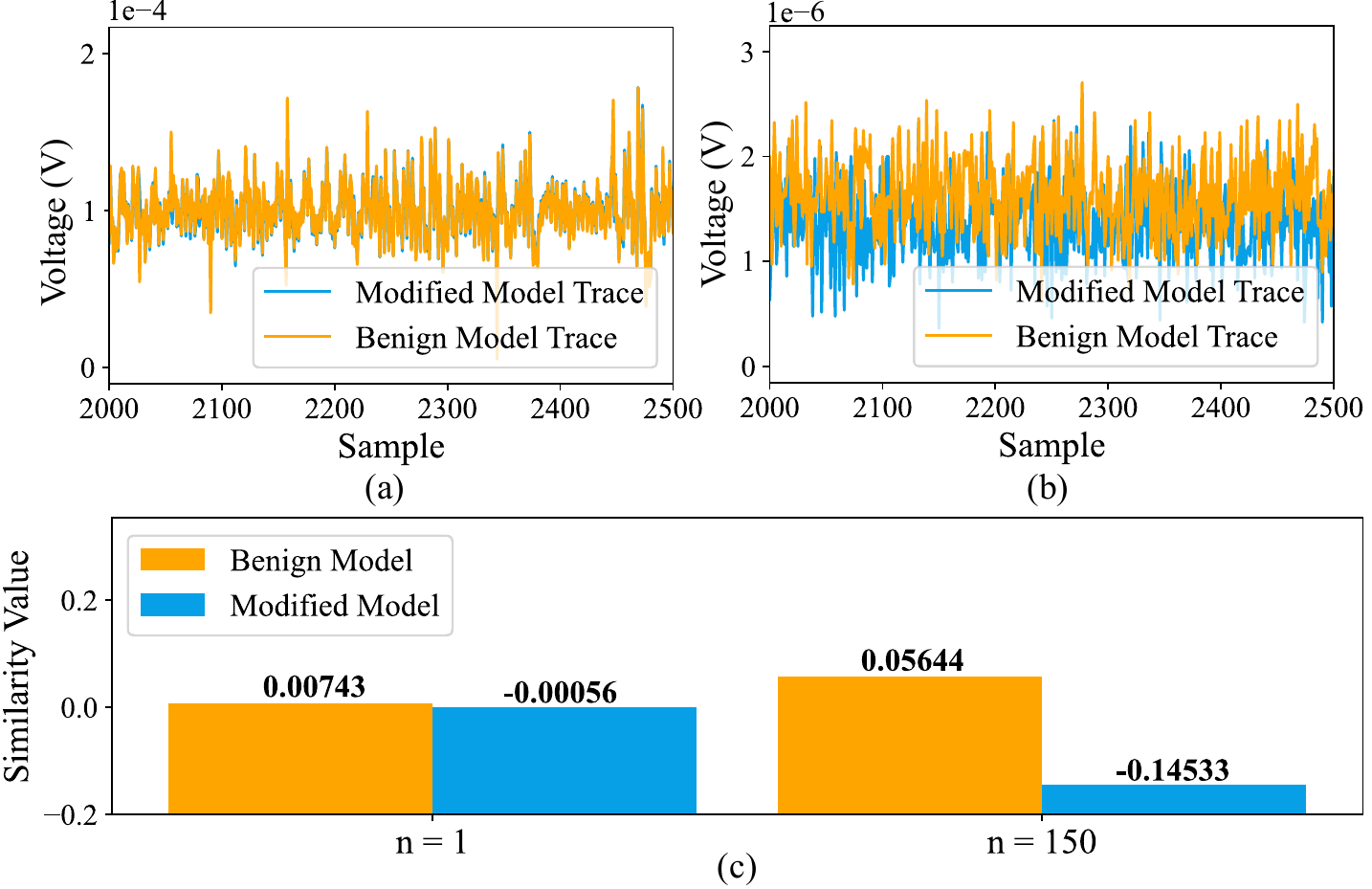}
  \caption{Voltage drop across shunt resistor during the final fully connected layer for benign/modified ResNet20. a) Single power trace overlaid for benign and modified models. b) Averaged power trace (n=150) overlaid for benign and modified models. c) Pearson correlation between benign and modified model traces to known benign trace.}
  
  \label{fig:motivation_traces_1}
\end{figure}

We have aggregated the resulting power traces for the final fully connected layer of both the benign and modified (poisoned) ResNet20 models in Fig. \ref{fig:motivation_traces_1}. To perform power-analysis-based integrity checking, we must reliably distinguish between benign and poisoned power traces. In Fig. \ref{fig:motivation_traces_1}(a), two test traces, benign and modified, are overlaid. Due to high noise levels, the traces are visually indistinguishable. Even when using Pearson correlation, a statistical similarity measurement, between a known benign trace and the two test traces, the difference between benign and modified traces is insignificant, as shown in Fig. \ref{fig:motivation_traces_1}(c). This indicates that distinguishing between benign and modified traces is non-trivial due to the presence of noise. To verify this assumption, we averaged 150 traces to reduce the noise floor and repeated the experiment. In this case, we observe visually distinguishable differences (Fig. \ref{fig:motivation_traces_1}(b)) as well as a significant gap in Pearson correlation (Fig. \ref{fig:motivation_traces_1}(c)) between the benign and modified models. This suggests that power analysis can indeed identify integrity violations in a TinyML model running on an MCU. However, to achieve this, we outline three primary challenges that must be addressed:

\begin{figure*}[t]
  \centering
  \includegraphics[width=\linewidth]{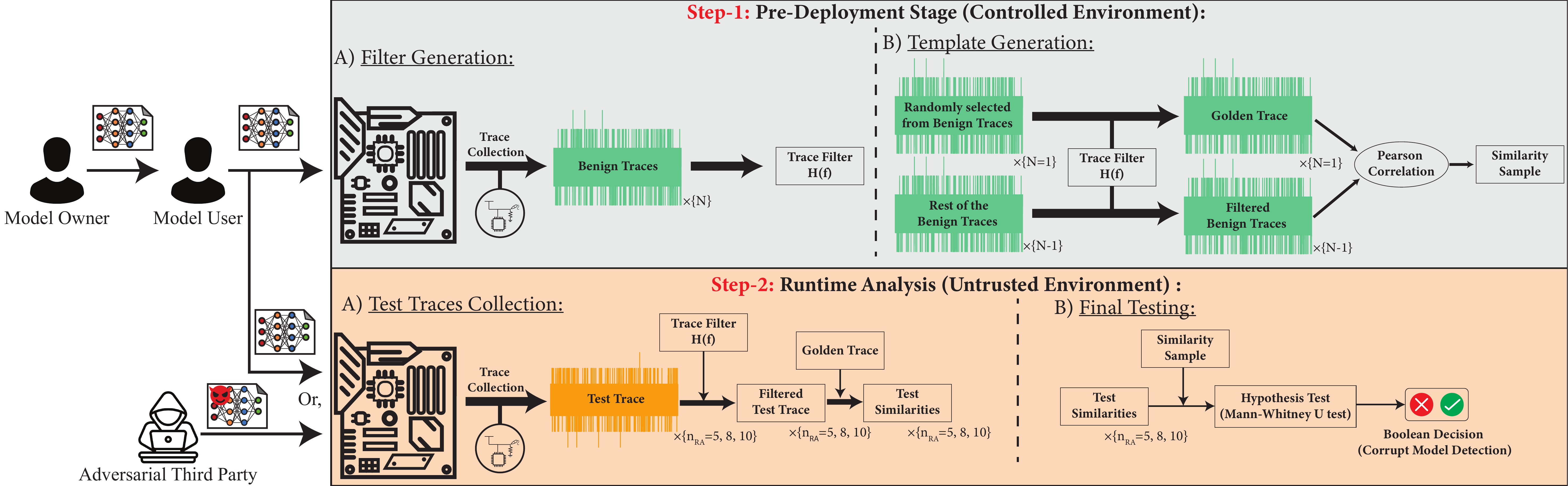}
  \caption{Overview of the proposed Michscan algorithm that uses power analysis for TinyML integrity checking.} 
  
  \label{fig:overall}
\end{figure*}

\begin{enumerate}
\label{sec:prob}

\item \textbf{Noise:} Fig. \ref{fig:motivation_traces_1}(a) shows that the high noise present in power traces makes integrity checking challenging. A method is needed to reduce noise without collecting many traces, which would impact performance.

\item \textbf{Template Generation:} Similarity measurements, such as Pearson correlation, provide a mathematically robust mechanism to differentiate between benign and modified traces. However, a baseline trace (i.e., a \textit{template} or \textit{signature}) is required for comparison.

\item \textbf{Quantifiable Authentication:} Fig. \ref{fig:motivation_traces_1}(c) shows that even two benign traces collected from the same MCU can be dissimilar (i.e., lower similarity values). This is due to both noise and natural variation in the device state (e.g., thermal effects or different peripherals running). A quantifiable method is required to distinguish between benign variations in power and integrity violations. 

\end{enumerate}

\section{Michscan Algorithm}

To address the challenges outlined in Sec. \ref{sec:prob}, we develop Michscan. Michscan is a runtime model authentication algorithm that uses power analysis to detect integrity violations in TinyML models. It operates by performing a hypothesis test (Mann-Whitney U-Test) to quantify the probability that the runtime power consumption of the MCU was produced by an un-modified model. By doing so, Michscan provides quantifiable integrity guarantees in real-time while operating under the black-box constraints commonly present in model licensing scenarios \cite{leroux2022tinymlops}. An overview of the MichScan algorithm is in Fig. \ref{fig:overall}. A corresponding flowchart for the requisite steps is in Fig. \ref{fig:enter-label}. We divide Michscan into 2 stages: Pre-Deployment and Runtime Analysis. Each of the stages with in-depth descriptions is introduced in turn. 


\subsection{Pre-Deployment Stage}
\label{sec:predeploy}


During the Pre-Deployment Stage of Michscan, we generate a \textit{golden template} along with a distribution of benign behavior (i.e., test sample) for the device. Both the \textit{golden template} and the \textit{test sample} are later used at runtime to estimate the probability of a measured trace originating from a benign model. We note that all power traces are generated for the same, randomly selected input to the model, which we refer to as the \textit{test input}. We note that the identity of the test input is irrelevant, only that it is the same for all collected power traces. The generated \textit{golden template} represents a benign scenario containing a power consumption pattern for the model running on the MCU under evaluation for the test input. Due to the hardware-specific nature of power consumption patterns, this pre-deployment stage must be performed per-device and after any modification to the model. An overview of the MichScan algorithm is depicted in the grey section of Fig. \ref{fig:overall}. A flowchart for the Michscan algorithm is in Fig. \ref{fig:enter-label}. 

To detect integrity violations, the \textit{golden template} must reflect the power consumption pattern of the intended model running on the device without modification. Hence, it must be collected during a period when the model is free of integrity violations. We operate under the assumption that the licensed model, immediately after purchase and prior to deployment, can be assumed to be free of any integrity violations (i.e., benign) if operated in a controlled environment. This assumption can be satisfied through traditional software supply-chain security strategies, such as audit trails and model signing, which ensure the model is in a benign state prior to deployment. We generate a set of power measurements for the known-benign model operating in the controlled setting. One of these power measurements is randomly selected to serve as a \textit{golden template} which is assumed to reflect the model running as intended. The remaining power traces are used to build a similarity distribution that captures the benign behavior of the device. We outline this procedure below. 

\subsubsection{\textbf{Template Generation}} \label{sec:noise} 

As demonstrated by the motivational example in Sec. \ref{sec:motivation}, noise is a primary impediment to differentiating benign from modified traces in power analysis. To detect integrity violations, it is crucial to isolate the components of a power trace produced by the control/data flow of the model running on the device from other signals. Ou et al. \cite{ou2016enhanced} noted that power/EM traces consist of three primary features: 1) a constant component, representing the baseline power consumption; 2) random variation or noise, an unwanted 
component caused by environmental factors or system fluctuations; and 3) systematic variation or signal, which changes based on the actual computations performed by the device. To effectively detect anomalies, one must focus on isolating this systematic variation from the constant component and random noise.

\begin{figure}
    \centering
    \includegraphics[width=1\linewidth]{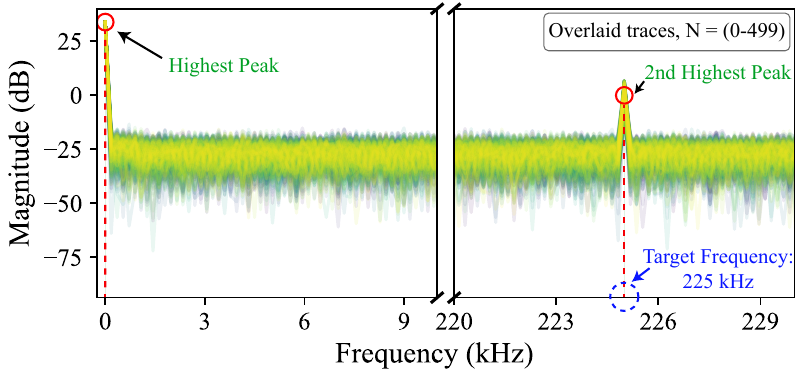}
    \caption{Frequency spectrum analysis of power traces collected during the pre-deployment stage. The overlaid FFT of N=500 traces shows two consistent peaks: the highest at 0 kHz (DC component) and a second peak at 225 kHz, with random noise appearing as irregular variations across the spectrum.}
    \label{fig:freq}
\end{figure}

To achieve this in Michscan, we collect a set of $N$ traces ($Tr=[tr_1,tr_2,\dots, tr_{N}]$) {Fig \ref{fig:enter-label} (Step 1)}. We then analyze all N=500 power traces in the frequency domain, shown in Fig. \ref{fig:freq}. Following Ou et al.'s observation \cite{ou2016enhanced}, power traces consist of three components: DC (constant), random noise, and systematic variation. For this device, frequency analysis consistently reveals two distinct peaks across all measurements: the highest peak at 0 Hz representing the constant DC component and a second peak at 225 kHz. For our particular setup, the presence of only a single distinct peak above the DC component indicates that this represents the systematic variation component of our system's operation as random noise would manifest as irregular peaks varying in both frequency and magnitude across trace measurements. However, we note that identifying systematic variation must be done per hardware design. Other hardware may require more careful analysis and exhibit more complex frequency representations. Based on this analysis, we consider the second-highest peak amplitude as our target frequency and create a Butterworth band-pass filter with a bandwidth of $\pm 1\%$ (Step 3). This filter is then applied to a single, randomly selected trace from the set of $N$ traces to create the \textit{golden template} (Step 5-6).

\begin{figure}[b!]
  \centering
  \includegraphics[width=\linewidth]{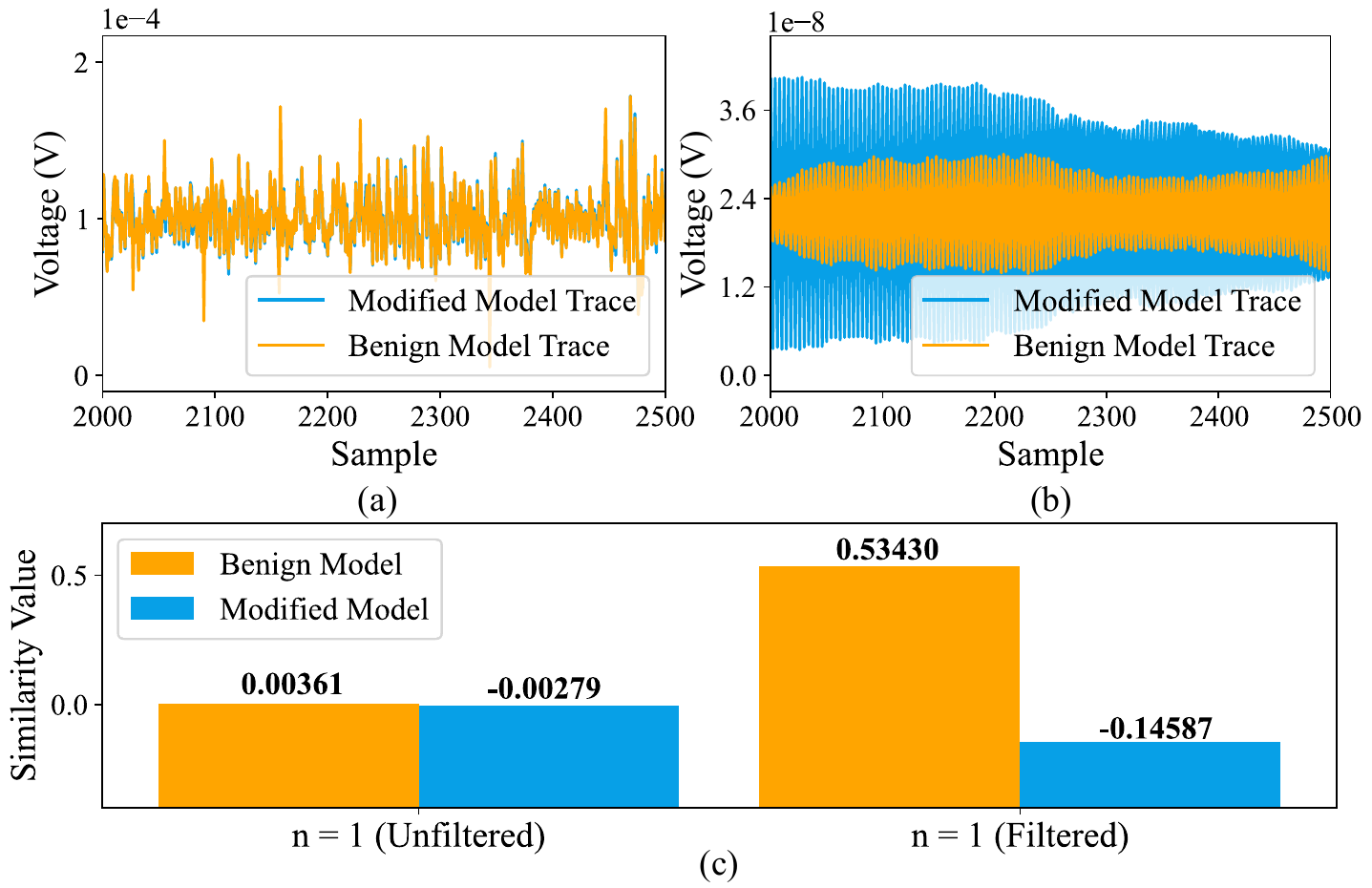}
  \caption{Voltage drop across shunt resistor during final fully connected layer for benign/modified ResNet20. a) Single power trace overlaid for benign/modified models. b) Filtered single power trace overlaid for benign/modified models. c) Pearson correlation between known-benign and benign/modified model trace for a single unfiltered and filtered power trace.}
  \label{fig:motivation_traces_2}
\end{figure}

We applied our Template Generation procedure to the power traces collected for the motivational example in Sec. \ref{sec:motivation}. Fig. \ref{fig:motivation_traces_2}(a) shows two visually indistinguishable power traces (benign and modified). Initially, Pearson correlation analysis between a benign trace and two test traces (benign and modified) yielded values near zero, indicating no discernible difference. However, Fig. \ref{fig:motivation_traces_2}(b) illustrates these traces after applying our filtering procedure, revealing clear visual distinctions between the benign and modified model traces. Moreover, post-filtering Pearson correlation analysis between the generated template and the two test traces demonstrated a significant dissimilarity. This demonstrates that our proposed filtering approach enables benign and modified traces to be differentiated, even when using only a single trace from each model.

\subsubsection{\textbf{Similarity Sample Generation}}
\label{sec:similarity}
As shown in Fig. \ref{fig:motivation_traces_2}, the generated template produces differentiable statistical similarity values (i.e., Pearson correlation) when applied to benign and modified traces. However, the \textit{golden template} and benign test traces still do not exhibit perfect similarity. This is caused by natural variations in noise and device state (e.g., different peripherals in use, temperature differences, noise, etc.). Depending on the device being used, this variation could be quite large. Therefore, a robust integrity-checking mechanism must reliably distinguish between natural, benign variations and those caused by integrity violations. To do so, we use the remaining $N-1$ known-benign traces, $\{Tr=[tr_1,tr_2,\dots, tr_{N-1}]\}$, collected during the Template Generation Phase (i.e., excluding the trace used as the \textit{golden template}) to serve as a set of samples from the population of benign traces. Each trace is filtered, as described in Sec. \ref{sec:noise}, and correlated with the \textit{golden template}. The resulting set of similarity values serves as a sample from the population of similarities produced by the device in a known benign state. We refer to this as the \textit{similarity sample}.  

During the Runtime Analysis phase, we use the \textit{similarity sample} to perform a Mann-Whitney U-Test, a non-parametric null hypothesis test \cite{nachar2008mann}. This test determines the probability (P-value) that the test sample, generated from traces collected at runtime, has the same median as the \textit{similarity sample}, the known-benign distribution. A low probability (e.g., P-value $< 10^{-5}$) of runtime-collected traces having the same median as the known-benign population indicates that the underlying distribution producing the test traces has shifted, suggesting that the model running on the device has been modified. This facilitates the quantifiable detection of integrity violations.

\subsection{Runtime Analysis Stage}
\label{sec: Runtime_need}

The Runtime Analysis Stage of Michscan periodically checks the integrity of the model running on the MCU. To do this, $n_{RA}$ number of power traces, which we refer to as \textit{test traces}, are collected for the \textit{test input}. This input must be the same randomly selected one from the Pre-Deployment Stage. These test traces are then filtered and correlated with the \textit{golden template}, as described in Sec. \ref{sec:noise}, to produce a set of similarity values. This set of values serves as a sample from the distribution of the Pearson correlations produced by the model currently running on the device. We refer to this set of samples as the \textit{test similarities}. We formulate integrity checking as a hypothesis testing problem and employ a Mann-Whitney U-Test for this purpose. The Mann-Whitney U-Test is a non-parametric hypothesis test that compares two independent samples (i.e., the test similarities and the \textit{similarity sample} to determine the probability that the \textit{test similarities} and the \textit{similarity sample} have an identical median. This approach is based on the fact that an integrity violation will impact the MCU's power consumption. As such, the underlying median of the \textit{test sample} is expected to shift, allowing detection through hypothesis testing. The Mann-Whitney U-Test calculates a P-value, which is the probability that the \textit{similarity sample} and the \textit{test similarities} are produced by a distribution with the same median. Importantly, users have the flexibility to set a threshold for the P-value that controls the sensitivity, allowing for the sensitivity of the detector to be tuned based on user application. Thus, the Mann-Whitney U-Test can quantify the probability that the underlying distribution producing \textit{test samples} has shifted, providing mathematically robust detection for integrity violations. To outline the approach, we begin by considering whether the necessary assumptions for a Mann-Whitney U-Test are satisfied. The Mann-Whitney U-Test has three necessary assumptions:

\begin{enumerate}
    \item \textit{Independence of Observations:} The \textit{test similarities} and the \textit{similarity sample} must be independent. Because each trace is collected from a separate model inference, the samples are independent, satisfying this assumption.
    \item \textit{Continuous Data:} The variable being compared between the \textit{test similarities} and \textit{similarity sample} must be continuous. Pearson correlation is a continuous variable that satisfies this assumption.
    \item \textit{Equal Shape and Spread:} The population producing the \textit{test similarities} and the \textit{similarity sample} must be similar in shape and spread. Both sets of samples are drawn from the same device running a similar application (i.e., a TinyML model), with the only variation expected to be the parameters and/or architecture of the model itself. Hence, the underlying distributions are assumed to have a similar shape and spread. 
\end{enumerate}

\begin{figure*}[ht]
    \centering
    \includegraphics[width=\linewidth]{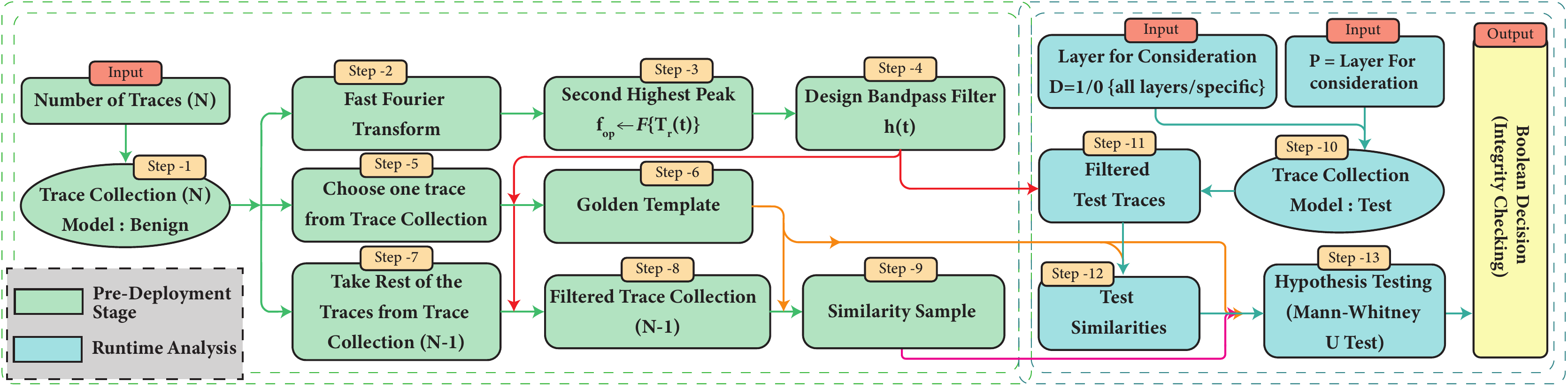}
    \caption{Flowchart for the Michscan algorithm separated into Pre-Deployment and Runtime Analysis stages. 
    } 
    \label{fig:enter-label}
\end{figure*}

Hence, the necessary conditions for using the Mann-Whitney U-Test are met. Now, we formulate the integrity checking problem being considered as a null hypothesis test. We define the null hypothesis $(H_0)$ based on a user-defined threshold $(P_{th})$ for the P-value (P) as follows.
\begin{enumerate}
    \item \textbf{Null Hypothesis ($H_0$):} The \textit{test similarities} and \textit{similarity sample} do not differ with statistical significance (i.e., $\text{P-value} \geq P_{th}$ that the median of the distribution producing both samples are the same).
    \item \textbf{Alternative Hypothesis ($H_1$):} The \textit{test similarities} and \textit{similarity sample} differ with statistical significance (i.e., $\text{P-value} < P_{th}$ that the median of the distribution producing both samples are significant difference).
\end{enumerate}

If the P-value ($P$) obtained from the Mann-Whitney U-Test is less than $P_{th}$, we reject the null hypothesis. The rejection of the null hypothesis indicates that the \textit{test similarities} are sampled from a distribution with a different median than the known-benign \textit{similarity sample}. \textbf{While Michscan flags this scenario as a potential integrity violation, it's important to note that this statistical deviation could stem from various sources beyond malicious modifications. Hardware component failures, environmental factors, or other system-level changes could also cause significant shifts in the underlying power consumption distribution.} This is a fundamental limitation of Michscan. While it can detect anomalous behavior with statistical confidence, it cannot definitively identify the source of the anomaly. However, in security-critical applications, any significant deviation from expected behavior warrants investigation, making this detection capability valuable regardless of the ultimate cause. Conversely, a probability $1-P_{th}$ of rejecting the null hypothesis when it is true (i.e., false positive) exists. $P_{th}$ can be tuned by the owner of the device to reflect the security requirements of the target application and device.

\subsubsection{\textbf{Runtime Analysis Methodology}} The right-side of Fig. \ref{fig:enter-label} contains a flow-chart outlining Michscans's Runtime Analysis algorithm. Runtime Analysis requires two inputs: the \textit{similarity sample} and the \textit{golden template} generated during Michscan's Pre-Deployment Stage (see Sec. \ref{sec:noise} and \ref{sec:similarity}). To proceed, a small set of ($n_{RA}$) \textit{test traces} 
are collected at runtime during inference while the pre-defined \textit{test input} is applied to the model (\textit{Step 10}). As we later show in Sec. \ref{sec:experiment}, these traces can characterize only a small subset (e.g., a single layer) of the model rather than the entire inference to reduce computational and memory overhead. Each of the $n_{RA}$ \textit{test traces} is then filtered using the bandpass filter generated in the Pre-Deployment Stage (see Sec. \ref{sec:noise}) and correlated with the \textit{golden template} (see Sec. \ref{sec:similarity}) (Step 11-12). This produces $n_{RA}$ similarity values to serve as the \textit{test similarities} (\textit{Step 12}). These serve as a set of samples from the population of similarity values produced by the current model running on the device, which may be modified. A Mann-Whitney U-Test is performed using the null hypothesis defined in Sec. \ref{sec: Runtime_need} to determine whether the \textit{test similarities} have been sampled from a population that differs with statistical significance from the \textit{similarity sample} (\textit{Step 13}) \cite{conroy2012hypotheses}. Based on the user-defined probability threshold ($P_{th}$), the null hypothesis is either accepted or rejected (\textit{Boolean Decision}). A rejection of the null hypothesis indicates that the population underlying the \textit{test similarities} differs with statistical significance from the \textit{similarity sample} due to the difference of the underlying medians. This indicates a statistically significant change in the device's power consumption, which we define as an integrity violation in the model or the device itself. Power traces for integrity checking can be collected during device downtime when no inference is being performed, or at periodic intervals determined by available resources and application requirements. To minimize overhead, these measurements can be collected by a small processor and analyzed remotely.

\subsection{Merits of the Michscan Methodology}

We make several observations on Michscan's merits.
\begin{itemize}
    \item \textit{Black-Box:} Because Michscan relies only on power analysis to detect possible integrity violations, no details of or modifications to the underlying model are required. This differs from prior work on integrity checking \cite{chen2019deepattest, chen2017targeted, stern2018emforced}, which requires either a cooperative model owner, access to model internals, or the model to be modified.
    \item \textit{Mathematically Quantifiable:} By adopting the Mann-Whitney U-Test, Michscan provides a quantifiable probability that the median of the \textit{similarity sample} has shifted from a distribution derived from a known benign device state. Because integrity violations in the underlying model alter the power profile of the device \cite{kocher1999differential, brier2004correlation}, this provides a mathematically robust detection mechanism for integrity checking.
    \item \textit{Tunable Sensitivity:} The Mann-Whitney U-Test produces a quantifiable probability for the underlying null hypothesis to be true. This allows a designer to both tune and mathematically quantify the sensitivity of the detector as well as the corresponding false positive rate to reflect the security requirements of a given device/application.
    \item \textit{Runtime:} Michscan requires a small number of periodic inferences in the model using a \textit{test input} in real-time with a low impact on the device performance.
    \item \textit{Generalized Integrity Checking:} Michscan makes no assumptions about the MCU, model, or underlying distribution of power traces. Hence, Michscan provides integrity checking for arbitrary systems in varied applications.
\end{itemize}

\section{Evaluation of Michscan Algorithm}
\label{sec:experiment}




To evaluate the Michscan algorithm, we implemented it on an STM32F303RC MCU containing an ARM Cortex-M4 core. As benchmark TinyML models for integrity checking, we selected 4 models from the MLPerf Tiny benchmark suite \cite{banbury2021mlperf}, providing a representative cross-section of model architectures, data modalities, and datasets. The details of each model are presented in Tab. \ref{table:models}. We collected power traces by measuring the voltage drop across a 10$\Omega$ shunt resistor using a 10-bit ADC at 96 MS/sec. Trigger signals were generated before and after each layer of the evaluation models to synchronize traces. While we explicitly generated these triggers, alternative approaches could be used to synchronize the traces. Since the test input ensures consistent model execution sequences, it can create recognizable power consumption patterns that can be used to predict starting points for a trace. This allows semi-automatic templating methods, such as \cite{trautmann2022semi}, to generate templates for synchronization. Alternatively, by triggering trace collections to when the test input is applied to the model, one can ensure that the model inference is captured for the test input. Depending on the hardware, this may be sufficient for analysis as is or can be further refined through templating procedures, such as \cite{trautmann2022semi}. However, as long as traces have been synchronized, the Michscan methodology does not fundamentally require fine-grain layer separation. 

\begin{table}[t]
    \caption{Overview of TinyML models used for evaluation.}
    \centering
    \scriptsize
    \label{table:models}
    \begin{tabular}{c c c c}
    \hline
    Use Case & Description & Dataset & Model \\ 
    \hline
        Keyword & Small vocabulary & Speech & \multirow{2}{*}{DS-CNN} \\
        Spotting & (keyword spotting) & Commands & \\
    \hline
        Visual Wake & Binary image & Visual Wake & \multirow{2}{*}{MobileNet} \\
        Words & classification & Words Dataset & \\
    \hline
        Image & Small image & Cifar10 & ResNet20 \\ \cline{3-4}
        Classification & classification &  Mnist & 2 Conv + 1 FC\\
    \hline
    \end{tabular}
    \label{tab:table_1}
\end{table}

To serve as integrity violations for evaluation, we considered one candidate threat from each family of integrity attacks outlined in Sec. \ref{sec:attack}: 
\begin{itemize}
\item \textit{Badnets \cite{gu2019badnets} (Trojan):} Badnets is a backdoor attack where the model is retrained to misclassify inputs containing adversarial triggers. Because the entire baseline benign model undergoes retraining, the resulting malicious model shares few parameters with the original benign model. To evaluate Michscan's performance, we generated 100 instances of each evaluated model subject to the Badnets attack. To do so, 20\% of the training data for a randomly-selected target class was infected with a random trigger sequence modifying 1\% of the input image/audio. The extensive modifications enable Michscan to detect the compromised models regardless of which layer it performs power analysis on, as the changes in model propagate and amplify through all layers.

\item \textit{Poisoning:} This attack fine-tunes a TinyML model with additional training data included in the dataset with incorrect, poisoned labels to degrade performance for a randomly selected class. To evaluate Michscan's performance, we generated 100 instances of each evaluated model subject to the poisoning attack. To do so, all layers but the final layer for each model were frozen, and a set of randomly-labeled data was used to fine-tune the model. All training hyperparameters for all models were untouched. This produces 100 malicious models with parameter modifications \textit{only} in the final layer compared to the benign model. This constitutes a difficult scenario as changes in the model occur \textit{only} in the final layer and do not accumulate throughout the entire model.

\item \textit{Bit-Flip Attack \cite{rakin2019bit} (Fault Injection):} The Bit-Flip Attack aims to degrade the performance of the model by flipping a small number of bits in the network's weight storage memory. For evaluation, we replicated the attack in a software scenario, where we generated 100 instances of each evaluated model with an integrity violation selected by \cite{rakin2019bit}. To generate each model, the Bit-Flip Attack was used to select and alter 4 parameters in the final layer. This presents a challenging scenario as the alteration is not amplified by subsequent layers.
\end{itemize}

Each unmodified (i.e., benign model) was run on the MCU for the Pre-Deployment Stage of Michscan (see Fig. \ref{sec:predeploy}). During this stage, a sample of 500 power traces was collected during benign model inference for a randomly selected \textit{test input} for each layer. Using the approach outlined in Sec. \ref{sec:noise}, the target frequency for the MCU was identified as 225kHz. A Butterworth filter, centered at 225kHz$\pm 1\%$, was designed to filter all $n = 500$ traces. One trace was randomly selected as the \textit{golden template}. The remaining 499 traces were correlated with the \textit{golden template} to be the \textit{similarity sample}.

For the Runtime Analysis Stage, we collected sets of $n_{RA} = {5,8,10}$ traces. These traces were then filtered and correlated with the \textit{golden template} to generate the \textit{test sample}. The resulting distribution, along with the \textit{test similarities}, undergoes Michscan's Runtime Analysis stage to determine whether a benign or modified model was running on the device. This process encompasses 16 evaluation scenarios (4 architectures with 3 possible model modifications and a benign, un-modified case). To evaluate Michscan, we analyze these results based on the four success criteria defined in Sec. \ref{sec:4_point}: Effectiveness, Efficiency, Reliability, and Generalizability.

\subsection{Success Criterion 1: Effectiveness}

\begin{figure*}[t]
  \centering
  \includegraphics[width=\linewidth]{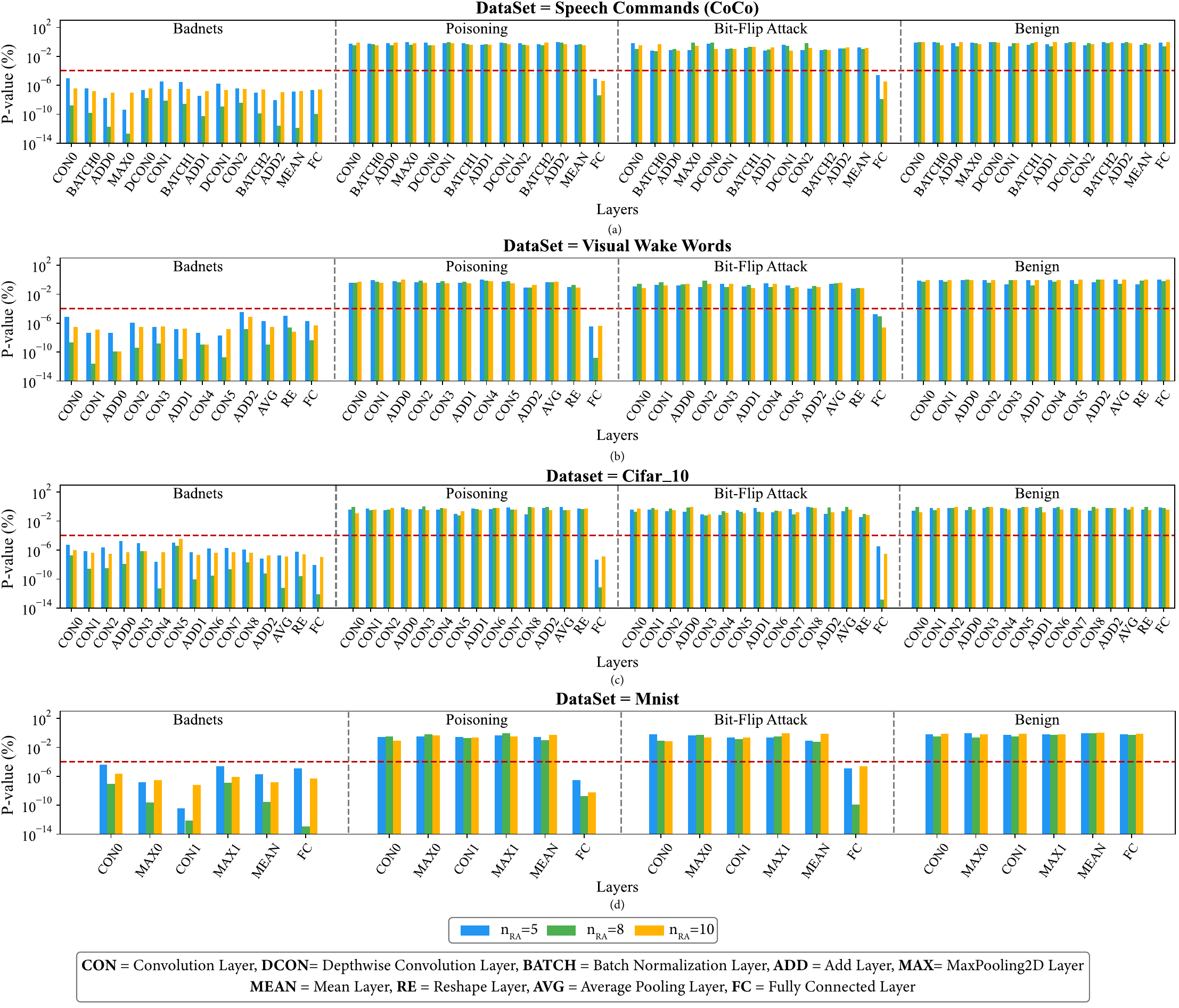} 
  \caption{Michscan P-values for each layer with test similarity size ($n_{RA}$) across evaluated TinyML models and integrity violations. }
  \label{fig:result}
\end{figure*}

The Effectiveness of Michscan corresponds to the likelihood of detecting integrity violations through the collected traces. Fig. \ref{fig:result} contains the average P-value from 100 instances of each test scenario and for three test similarity group sizes ($n_{RA} = 5,8,10$). The user-defined detection threshold, visualized as a dashed red line in the figure, was defined as $P_{th} = 10^{-5}$. A detection will occur when the P-value (P) is lower than $P_{th}$ (i.e., $P<10^{-5}\%$). In this case, the null hypothesis is rejected, indicating that the median for the distribution producing \textit{test similarities} is statistically different from the distribution producing the similarity sample. 
Tab. \ref{table:results2} contains the number of correct classifications of each test case out of the 100 trials for ($n_{RA}=3, 5, 10$) where only the final fully connected layer is considered. As discussed in Sec. \ref{sec:mann}, the Mann-Whitney U-test suggests $n_{RA} \geq 5$ for statistical validity \cite{nachar2008mann}. Our results support this requirement: for distributions with $n_{RA}= 5$ and $10$, Michscan achieved 100\% detection of all integrity violations across all three attack types through the final fully connected layer with no false positives in benign scenarios. In contrast, $n_{RA} = 3$ showed significantly reduced detection rates and increased false positives. We note that this performance degradation is consistent across networks, but does appear to vary by the type of integrity violation. For example, bit-flip attacks, which make the slowest severity changes have the highest false negative rate. As noted in \cite{nachar2008mann}, a larger test similarity makes the U-Test more noise-resilient because there are more rank combinations to calculate over. Hence, with a higher noise, a larger value of $n_{RA}$ may be required. This provides a trade-off between efficiency and reliability. However, for $n_{RA} \geq 5$, Michscan achieved 100\% effectiveness, detecting all 2400 integrity violations and correctly classifying all 800 benign cases.

\begin{figure}[t]
    \centering
    \includegraphics[width=1\linewidth]{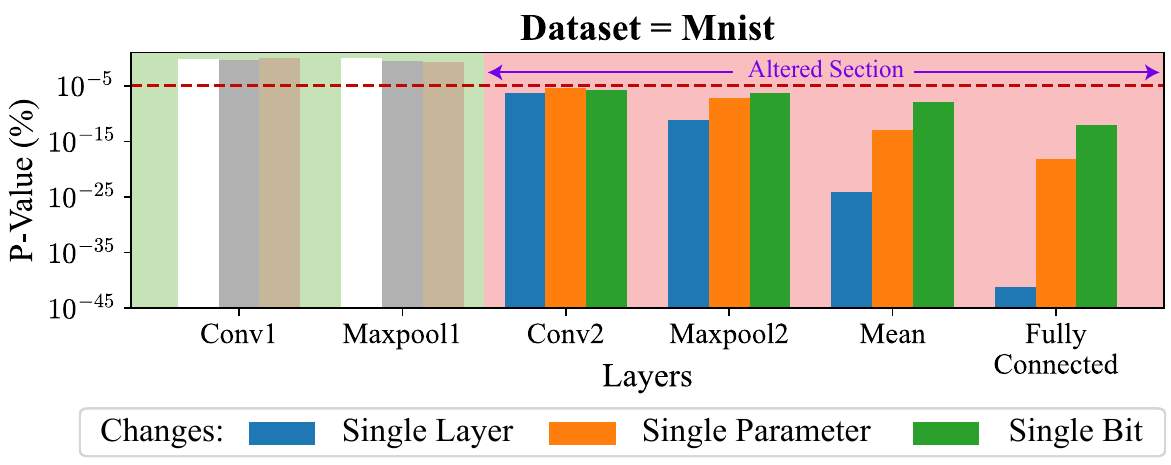}
    \caption{Layer-wise P-values for different integrity violations introduced in Conv2 with test similarity size of $n_{RA}=5$.}
    \label{fig:layer_conv2}
\end{figure}

Additionally, to assess the severity of integrity violation detected by Michscan, we introduced integrity violations impacting a different number of parameters (i.e., with different severity) into the Conv2 layer of the Mnist model. Specifically, 100 instances of integrity violations were randomly generated of 3 severity levels, namely modifying the parameters in 1) a single layer, 2) a single parameter, and 3) a single bit. These results are aggregated in Fig. \ref{fig:layer_conv2}. We note that Michscan successfully detected all 300 integrity violations in Conv2 and all following layers, regardless of their severity. Based on this we draw 3 conclusions. First, these results indicate that modifications in a model propagate through and are amplified as the progress deeper into the network. When changes are introduced early (e.g., Conv2) in the MNIST model, their effects cascade through the network, resulting in progressively lower P-values deeper in the architecture. This propagation pattern remains consistent across different modification types: parameter changes occurring in a single layer, parameter, or bit. 
Second, Michscan can classify the severity of an integrity violation in a model through the resulting P-value. For example, for Michscan deployed in the fully connect layer, full-layer modifications produce exponentially lower P-values compared to single parameter changes, which in turn show exponentially lower P-values than single-bit modifications. Finally, we note that while Michscan cannot distinguish between architectural and parameter modifications, it is capable of detecting both types of modifications since any change to the model produces detectable shifts in the device's power consumption pattern. This empirical analysis demonstrates Michscan's ability to detect violations both at their source and through their propagated effects throughout the network.

We draw two main conclusions from these results. First, since model integrity violations must always impact the final layer of a model to affect its output, Michscan can reliably detect integrity violations by monitoring only the final layer (Fig. \ref{fig:result} and Tab. \ref{table:results2}). This can effectively reduce Michscan's computational footprint. Second, these results indicate that Michscan exhibits strong detection capabilities, demonstrating high effectiveness (Fig. \ref{fig:result}, Tab. \ref{table:results2}, and Fig. \ref{fig:layer_conv2}).

\begin{table}[b]
    \centering
    \scriptsize
    \caption{Michscan detection performance across test similarity size ($n_{RA}=3,5,10$). All results are for Michscan applied to the final fully connected layer.}
    \begin{tabular}{c c c c c c}
    \hline
     Dataset &$n_{RA}$ & Badnets & Poisoning &\makecell{Bit-Flip\\Attack } &\makecell{None\\(Benign)}\\
    \hline
    \multirow{3}{*}{\makecell{Keyword\\Spotting}} 
        & 3 & \cellcolor[HTML]{F9BEBF}78/100 & \cellcolor[HTML]{F9BEBF}43/100 & \cellcolor[HTML]{F9BEBF}16/100 & \cellcolor[HTML]{F9BEBF}23/100 \\
        & 5 & \cellcolor[HTML]{C6E2B7}100/100 & \cellcolor[HTML]{C6E2B7}100/100 & \cellcolor[HTML]{C6E2B7}100/100 & \cellcolor[HTML]{C6E2B7}100/100 \\
        & 10 & \cellcolor[HTML]{C6E2B7}100/100 & \cellcolor[HTML]{C6E2B7}100/100 & \cellcolor[HTML]{C6E2B7}100/100 & \cellcolor[HTML]{C6E2B7}100/100 \\ \hline
    \multirow{3}{*}{\makecell{Visual Wake\\Words (COCO)}} 
        & 3 & \cellcolor[HTML]{F9BEBF}61/100 & \cellcolor[HTML]{F9BEBF}52/100 & \cellcolor[HTML]{F9BEBF}27/100 & \cellcolor[HTML]{F9BEBF}52/100 \\
        & 5 & \cellcolor[HTML]{C6E2B7}100/100 & \cellcolor[HTML]{C6E2B7}100/100 & \cellcolor[HTML]{C6E2B7}100/100 & \cellcolor[HTML]{C6E2B7}100/100 \\
        & 10 & \cellcolor[HTML]{C6E2B7}100/100 & \cellcolor[HTML]{C6E2B7}100/100 & \cellcolor[HTML]{C6E2B7}100/100 & \cellcolor[HTML]{C6E2B7}100/100 \\ \hline   
    \multirow{3}{*}{\makecell{Cifar-10}} 
        & 3 & \cellcolor[HTML]{F9BEBF}84/100 & \cellcolor[HTML]{F9BEBF}77/100 & \cellcolor[HTML]{F9BEBF}15/100 & \cellcolor[HTML]{F9BEBF}46/100 \\
        & 5 & \cellcolor[HTML]{C6E2B7}100/100 & \cellcolor[HTML]{C6E2B7}100/100 & \cellcolor[HTML]{C6E2B7}100/100 & \cellcolor[HTML]{C6E2B7}100/100 \\
        & 10 & \cellcolor[HTML]{C6E2B7}100/100 & \cellcolor[HTML]{C6E2B7}100/100 & \cellcolor[HTML]{C6E2B7}100/100 & \cellcolor[HTML]{C6E2B7}100/100 \\ \hline
    \multirow{3}{*}{\makecell{Mnist}} 
        & 3 & \cellcolor[HTML]{F9BEBF}59/100 & \cellcolor[HTML]{F9BEBF}47/100 & \cellcolor[HTML]{F9BEBF}09/100 & \cellcolor[HTML]{F9BEBF}74/100 \\
        & 5 & \cellcolor[HTML]{C6E2B7}100/100 & \cellcolor[HTML]{C6E2B7}100/100 & \cellcolor[HTML]{C6E2B7}100/100 & \cellcolor[HTML]{C6E2B7}100/100 \\
        & 10 & \cellcolor[HTML]{C6E2B7}100/100 & \cellcolor[HTML]{C6E2B7}100/100 & \cellcolor[HTML]{C6E2B7}100/100 & \cellcolor[HTML]{C6E2B7}100/100 \\
    \hline
    \end{tabular}
    \label{table:results2}
\end{table}

\subsection{Success Criterion 2: Efficiency} 
The Efficiency of Michscan corresponds to the number of test cases required to detect an integrity violation reliably. Ideally, this should be minimized to limit Michscan's impact on device performance. Table \ref{table:results2} shows that all 2400 evaluated integrity violations and 800 benign cases were properly classified for test similarity distribution size $(n_{RA}=5$ and $10)$. Hence, based on these results, Michscan can reliably detect integrity violations within 5 measurements of model inference, indicating a high degree of efficiency. Additionally, we note that the user can set the interval at which Michscan checks model integrity, allowing Michscan's efficiency to be tuned.

\subsection{Success Criterion 3: Reliability} 

The Reliability of Michscan corresponds to the rate of false positives (i.e., an unmodified model detected as modified) and false negatives (i.e., a modified model not detected as such). High reliability is characterized by low rates of both false positives (FP) and false negatives (FN). Tab. \ref{table:results2} presents the results of all 3200 scenarios evaluated by Michscan for $n_{RA}= (5$ and $10$) across benign and various attack scenarios observed for the final fully connected layer. As shown in Tab. \ref{table:results2}, no FPs or FNs were observed. This indicates that Michscan exhibits high reliability across the diverse range of integrity violations and model architectures that were considered.

\subsection{Success Criterion 4: Generalizability} The Generalizability of Michscan corresponds to its ability to function across various TinyML models and integrity violations. The results in Fig. \ref{fig:result} demonstrate that Michscan detected all 3 candidate integrity violation instances across 4 TinyML models with varying architectures and data modalities. This indicates that Michscan generalizes well, suggesting its applicability to a wide range of integrity-critical applications.


\section{Conclusion}

This work introduces Michscan, a runtime integrity-checking mechanism for TinyML models. Michscan ensures integrity at runtime by monitoring the instantaneous power consumption of an MCU executing a TinyML model. Unlike prior work, Michscan operates in a black-box environment, requiring no knowledge of the TinyML model running on the device or cooperation by the model's owner. Such constraints may arise in a model licensing scenario. To detect potential integrity violations, Michscan utilizes a non-parametric statistical hypothesis test (Mann-Whitney U-Test) to quantify the underlying power consumption of the device has changed, indicating an integrity violation. To evaluate Michscan, we implemented it in an STM32F303RC MCU with an ARM Cortex-M4 running 4 TinyML models in the presence of 3 model integrity violations. Michscan successfully detected all integrity violations at runtime, even those modifying a few model parameters, using power data from 5 inferences. All detected integrity violations had a negligible probability of ($P < 10^{-5}\%$) for being produced from an unmodified model.

\section*{Acknowledgments}
This material is based upon work supported by the National Science Foundation under Grant No. 2245573.

\bibliographystyle{IEEEtran}
\bibliography{New_Reference}

\end{document}